\documentclass[showpacs,aps,prb,twocolumn,superscriptaddress,10pt, nofootinbib]{revtex4-2}
\usepackage{graphicx} 
\usepackage{bm}
\usepackage{color}
\usepackage{xcolor}
\usepackage{amsmath}
\usepackage{amssymb}
\usepackage{enumerate}
\usepackage[bottom]{footmisc}
\usepackage{xspace}
\usepackage{mathrsfs}
\usepackage{mathptmx}
\usepackage{gensymb}
\usepackage[colorlinks=true,linkcolor=blue,citecolor=blue,urlcolor=blue]{hyperref}
\usepackage{orcidlink}
\usepackage[utf8]{inputenc}
\usepackage[normalem]{ulem}
\usepackage{marvosym}

\newcommand\blfootnote[1]{%
  \begingroup
  \renewcommand\thefootnote{}\footnote{#1}%
  \addtocounter{footnote}{-1}%
  \endgroup
}

\begin{document}

\title{Spin Hall magnetoresistance at the altermagnetic insulator/Pt interface}
\author{Miina~Leiviskä$^{a,\star}$}
\affiliation{Institute of Physics, Czech Academy of Sciences, Prague, Czechia}
\author{Reza Firouzmandi$^a$}
\affiliation{Institut f\"ur Festk\"orperforschung, Leibniz IFW Dresden, 01069 Dresden, Germany}
\author{Kyo-Hoon Ahn}
\affiliation{Institute of Physics, Czech Academy of Sciences, Prague, Czechia}
\author{Peter Kuba\v s\v cik}
\affiliation{Faculty of Mathematics and Physics, Charles University, Prague, Czechia}
\author{Zbynek Soban}
\affiliation{Institute of Physics, Czech Academy of Sciences, Prague, Czechia}
\author{Satya Prakash Bommanaboyena}
\affiliation{Institute of Physics, Czech Academy of Sciences, Prague, Czechia}
\author{Christoph M\"{u}ller}
\affiliation{Institute of Physics, Czech Academy of Sciences, Prague, Czechia}
\affiliation{Faculty of Mathematics and Physics, Charles University, Prague, Czechia}
\author{Dominik Kriegner}
\affiliation{Institute of Physics, Czech Academy of Sciences, Prague, Czechia}
\author{Sebastian Sailler}
\affiliation{Department of Physics, University of Konstanz, Konstanz, Germany}
\author{Denise Reustlen}
\affiliation{Department of Physics, University of Konstanz, Konstanz, Germany}
\author{Michaela Lammel}
\affiliation{Department of Physics, University of Konstanz, Konstanz, Germany}
\author{Kranthi Kumar Bestha}
\affiliation{Institut f\"ur Festk\"orperforschung, Leibniz IFW Dresden, 01069 Dresden, Germany}
\affiliation{Institute of Solid State and Materials Physics, TU Dresden, Dresden, Germany}
\author{Mat\v{e}j H\'yvl}
\affiliation{Institute of Physics, Czech Academy of Sciences, Prague, Czechia}
\author{Libor \v{S}mejkal}
\affiliation{Max Planck Institute for the Physics of Complex Systems, Dresden, Germany}
\affiliation{Max Planck Institute for Chemical Physics of Solids, Dresden, Germany}
\affiliation{Institute of Physics, Czech Academy of Sciences, Prague, Czechia}
\author{Jakub \v{Z}elezn\'{y}}
\affiliation{Institute of Physics, Czech Academy of Sciences, Prague, Czechia}
\author{Anja U. B. Wolter}
\affiliation{Institut f\"ur Festk\"orperforschung, Leibniz IFW Dresden, 01069 Dresden, Germany}
\author{Monika Scheufele}
\affiliation{Walther-Meißner-Institut, Bayerische Akademie der Wissenschaften, Garching, Germany}
\affiliation{Technical University of Munich, TUM School of Natural Sciences, Physics Department, Garching, Germany}
\author{Johanna Fischer}
\affiliation{Université Grenoble Alpes, CEA, CNRS, Spintec, Grenoble, France}
\author{Matthias Opel}
\affiliation{Walther-Meißner-Institut, Bayerische Akademie der Wissenschaften, Garching, Germany}
\author{Stephan Geprägs}
\affiliation{Walther-Meißner-Institut, Bayerische Akademie der Wissenschaften, Garching, Germany}
\author{Matthias Althammer}
\affiliation{Walther-Meißner-Institut, Bayerische Akademie der Wissenschaften, Garching, Germany}
\author{Bernd B\"{u}chner}
\affiliation{Institut f\"ur Festk\"orperforschung, Leibniz IFW Dresden, 01069 Dresden, Germany}
\affiliation{Institute of Solid State and Materials Physics and W\"urzburg-Dresden Cluster of Excellence ct.qmat, Technische Universität Dresden, 01062 Dresden, Germany}
\affiliation{Center for Transport and Devices, Technische Universität Dresden, 01069 Dresden, Germany}
\author{Tomas Jungwirth}
\affiliation{Institute of Physics, Czech Academy of Sciences, Prague, Czechia}
\affiliation{School of Physics and Astronomy, University of Nottingham, Nottingham, UK}
\author{Luk\'a\v{s} N\'advorn\'ik}
\affiliation{Faculty of Mathematics and Physics, Charles University, Prague, Czechia}
\author{Sebastian T. B. Goennenwein}
\affiliation{Department of Physics, University of Konstanz, Konstanz, Germany}
\author{Vilmos Kocsis}
\affiliation{Institut f\"ur Festk\"orperforschung, Leibniz IFW Dresden, 01069 Dresden, Germany}
\author{Helena Reichlova}
\affiliation{Institute of Physics, Czech Academy of Sciences, Prague, Czechia}
\begin{abstract}
The resistance of a heavy metal can be modulated by an adjacent magnetic material through the combined effects of the spin Hall effect, inverse spin Hall effect, and dissipation of the spin accumulation at the interface. This phenomenon is known as the spin Hall magnetoresistance. The dissipation of the spin accumulation can occur via various mechanisms, with spin-transfer torque being the most extensively studied. In this work, we report the observation of spin Hall magnetoresistance at the interface between platinum and an insulating altermagnetic candidate, Ba$_2$CoGe$_2$O$_7$. Our findings reveal that this heterostructure exhibits a relatively large spin Hall magnetoresistance signal, which is anisotropic with respect to the crystal orientation of the current channel. We explore and rule out several potential explanations for this anisotropy and propose that our results may be understood in the context of the anisotropies of the spin current channels across the Pt/altermagnetic Ba$_2$CoGe$_2$O$_7$ interface. 

\end{abstract}

\maketitle
\section{Introduction}

\begin{figure}[ht!]
\includegraphics[width=0.40\textwidth, trim=0 0 0 0]{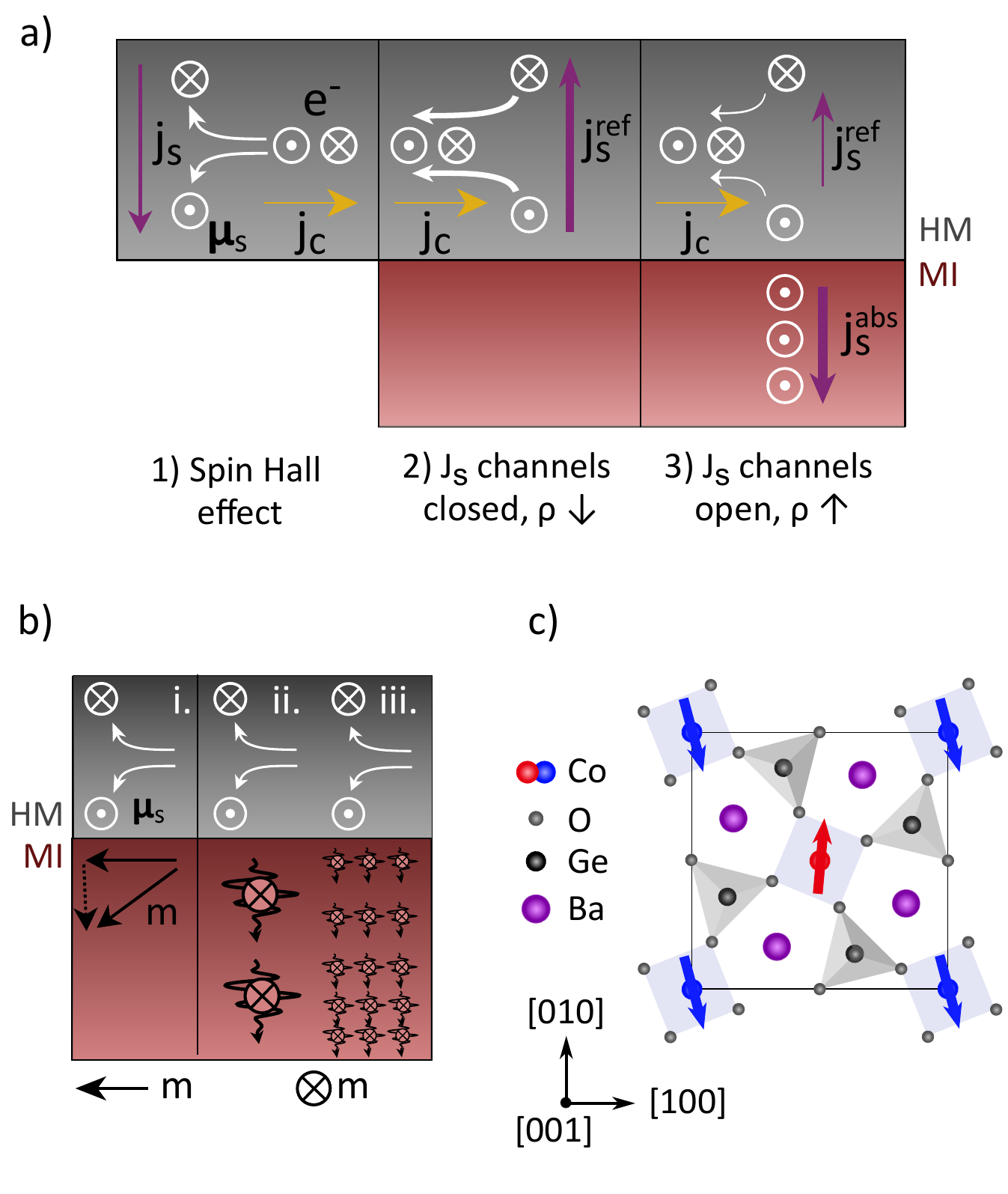}
\caption{(Color Online) a)  Spin Hall magnetoresistance in HM/MI heterostructures, where charge current ($\mathbf{j_c}$) generates a spin current (1) that can be either reflected ($\mathbf{j^{ref}_s}$) or absorbed ($\mathbf{j^{abs}_s}$) at the interface depending on the available spin current channels across the interface (2,3). When there are no available channels the spin current is reflected and converted back to charge current (2) and when the channels are open the spin current is absorbed by the MI (3). This modulates the resistivity of the HM. b) Three channels contributing to the SMR ratio in magnetic insulators in the dc-limit: i.  STT and spin pumping, ii. excitation of longitudinal magnons, and iii. magnon capacitance. Channel i. is relevant when $\pmb{\mu}_s\perp\mathbf{m}$ and channels ii. and iii. are relevant when $\pmb{\mu}_s\parallel\mathbf{m}$. c) The crystal structure of BCGO. The spins of the magnetic sublattices are marked in red and blue and the spin canting is exaggerated for the sake of clarity.}
\label{fig:1}
\end{figure}

Spin Hall magnetoresistance (SMR) is an actively researched magnetoresistive effect present in bilayers of a magnet - here insulating (MI) - and a heavy metal (HM) \cite{Nakayama2013, Chen2013, Althammer2013}. It relies on the concerted action of spin Hall (SHE) and inverse spin Hall effects (ISHE): driving a current ($\mathbf{j_c}$) in the HM layer generates a spin accumulation ($\pmb{\mu}_s$) at the HM$\vert$MI interface via SHE and depending on the spin transparency of this interface the generated spin current ($\mathbf{j_s}$) can flow into the MI ($\mathbf{j^{abs}_s}$) while any back-reflected spin current ($\mathbf{j^{ref}_s}$) is re-converted to a charge current via ISHE. This process is illustrated in Figure \ref{fig:1}a. The spin current transparency of the interface is sensitive to the (sublattice) magnetic moment direction ($\mathbf{m}$) of the MI layer - rotating $\mathbf{m}$ will modulate the resistivity of the HM layer. Typically, when $\pmb{\mu}_s$ and $\mathbf{m}$ are parallel more spin current is reflected back at the HM$\vert$MI interface resulting in a low-resistance state, while when $\pmb{\mu}_s$ and $\mathbf{m}$ are perpendicular to each other, more spin current is absorbed by the MI resulting in a high-resistance state. The SMR ratio, which quantifies the relative resistance difference between these two states, is typically of the order of 0.01\% in MI/Pt heterostructures \cite{Althammer2013, Isasa2014, Han2014, Putter2017, Dong2017, Hoogeboom2017, Fischer2018, Fischer2020}.

\blfootnote{$^a$ These authors contributed equally. *leiviska@fzu.cz
}

In the initial theoretical models describing the SMR, the variation of the interface spin-current transparency as a function of the relative angle of $\pmb{\mu}_s$ and $\mathbf{m}$ was explained in terms of the spin-transfer torque (STT) at the interface \cite{Chen2013}: if $\pmb{\mu}_s$ and $\mathbf{m}$ are parallel, STT is zero and the spin current is reflected back while if $\pmb{\mu}_s$ and $\mathbf{m}$ are perpendicular, maximum STT is exerted on $\mathbf{m}$ and the spin current is absorbed by the MI. However, in principle any mechanism through which the spin accumulation can dissipate from the interface will contribute to the SMR ratio. For example, thermal fluctuations allow for spin-flip scattering at the interface and give rise to injection/annihilation of thermal magnons, resulting in a maximum flow of spin current when $\mathbf{m}\parallel\pmb{\mu}_s$ \cite{Cornelissen2015, Goennenwein2015, Velez2016, Zhang2019, Kato2020}. A recent theoretical work by Reiss \textit{et al.} provides a comprehensive account of the various channels through which spin current can enter the MI layer \cite{Reiss2021}. In dc-measurements on HM$\vert$MI heterostructures, three primary contributions (summarized in Figure \ref{fig:1}b) are expected: i) STT and spin-pumping at the HM$\vert$MI interface, ii) incoherent thermal magnon creation/annihilation at the HM$\vert$MI interface, and iii) magnon capacitance (i.e. accumulation and dissipation of magnons in the MI).

Altermagnets are a newly identified class of collinear compensated magnets which differ from antiferromagnets by their combined spin and crystal symmetry. Altermagnets have characteristic spin-degenerate nodes and alternating even-parity spin polarization splitting that breaks time-reversal symmetry \cite{Smejkal2022a, Smejkal2022b, Jungwirth2024, Jungwirth2024b}. They exhibit unconventional properties that are potentially favorable for spintronics and magnonics, such as the unconventional anomalous Hall effect \cite{Smejkal2020, Mazin2021}, the generation of highly anisotropic and strongly spin-polarized currents in the absence of a net magnetization \cite{GonzalezHernandez2021, Bose2022, Karube2022, Bai2023}, the accompanying unconventional giant/tunneling magnetoresistance \cite{Smejkal2022gmr, Samanta2024} and spin-torque effects \cite{ Karube2022, Bai2023}, or anisotropic chirality-split magnon dispersions in the absence of external magnetic field \cite{Smejkal2023, Cui2023}. Notably, also the record SMR ratio (0.25 \%) in insulators has been observed in heterostructures of Pt and the insulating altermagnetic candidate $\alpha$-Fe$_2$O$_3$ \cite{Fischer2020}. This greatly surpasses the SMR ratio in heterostructures incorporating ferrimagnetic insulators, such as yttrium ion garnet (0.16 \%) \cite{Althammer2013}, and antiferromagnetic insulators, such as NiO (0.08 \%) \cite{Fischer2018}. This may suggest that the type of magnetic ordering of the MI plays a role in determining the SMR ratio. Recently, SMR has been even suggested as a tool for determining the spin order in altermagnets \cite{Kobayashi2024}.

In this work, we present our measurements of SMR in an altermagnetic candidate and ferroelectric insulator Ba$_2$CoGe$_2$O$_7$ (BCGO) \cite{Smejkal2024}. We first detail the fabrication of devices for SMR experiments on bulk single crystals of BCGO, followed by the analysis of the SMR features on BCGO/Pt heterostructures. Notably, we show that the SMR ratio (of the order of 2$\times10^{-4}$) is anisotropic depending on the crystal direction of the applied current. We discuss this observation in the context of various factors such as magnetic domains, magnetocrystalline anisotropy, electric polarization, and the non-relativistic and relativistic symmetries of BCGO. Finally, we suggest mechanisms including anisotropic spin mixing conductance and anisotropic magnon dispersion, which could give rise to the observed anisotropy. Overall, our results demonstrate a robust and anisotropic SMR response in an altermagnetic candidate and multiferroic material with a well-known spin structure \cite{Hutanu2012}. We thereby add another candidate material to the pool for altermagnetic spin transport research while driving forward the research towards a tunable and electrically controllable SMR response.

\section{SMR in B\MakeLowercase{a}$_2$C\MakeLowercase{o}G\MakeLowercase{e}$_2$O$_7$/P\MakeLowercase{t} heterostructures}

BCGO is an insulating altermagnetic candidate material that crystallizes in a non-centrosymmetric tetragonal structure in space group $P\bar{4}2_1m$ (Figure \ref{fig:1}c) \cite{Hutanu2011,Hutanu2012} and belongs to the spin-point group $^1$4$^2$m$^2$m \cite{Smejkal2024}. Below the critical temperature of 6.7 K, the sublattice magnetic moments residing on the Co$^{2+}$ ions adopt an antiparallel configuration in the (001) plane \cite{Hutanu2012}. The N\'eel vector is defined as $\mathbf{l}=\frac{\mathbf{m_1-m_2}}{2}$, where $\mathbf{m_{1,2}}=\mathbf{M_{1,2}/|M_{1,2}|}$ and $\mathbf{M_{1,2}}$ are the sublattice magnetizations. A finite DMI-vector along the [001] direction allows for the slight in-plane canting of the sublattice moments and thus a weak net magnetization ($\mathbf{M}=\mathbf{M_1+M_2}$), the extent of which depends on the orientation of the sublattice moments but is of the order of 0.01 $\mu_B/$f.u. \cite{Sato2003, Yi2008, Solovyev2015, Thoma2022}. BCGO is also ferroelectric below the critical temperature of 6.7 K - the finite polarization has been attributed to the p-d hybridization between the transition metal and the ligand \cite{Murakawa2010, Solovyev2015}. Upon the in-plane rotation of the sublattice magnetic moments, $P_x$ and $P_y$ are zero while $P_z$ shows a cosine dependency with maximum $P_z$ when the magnetic field is parallel to $\langle 110\rangle$ \cite{Murakawa2010, Solovyev2015}. When a strong out-of-plane field (order of 5 T) is applied, the spins cant away from the (001)-plane, which results in the $P_z$ decreasing and $P_x$ and $P_y$ becoming finite \cite{Murakawa2010, Solovyev2015}. 

The single crystal of BCGO was grown by the floating zone method in the presence of artificial air (80\,$\%$ N$_2$ and 20\,$\%$ O$_2$). The [001] normal surfaces were polished using Al$_2$O$_3$ lapping films (261X, 3M) up to 0.3\,$\mu$m surface quality. The final surface quality of $\sim$ 5 nm was realized with chemical-mechanical polishing using 0.02\,$\mu$m silica suspension (MasterMet2, Buehler). The surface quality of the BCGO was studied with atomic force microscopy on Bruker ICON machine using the Aspire CFM probe with conical tip and tip radius $<$10 nm.
\begin{figure}[ht!]
\includegraphics[width=0.45\textwidth, trim=0 0 0 0]{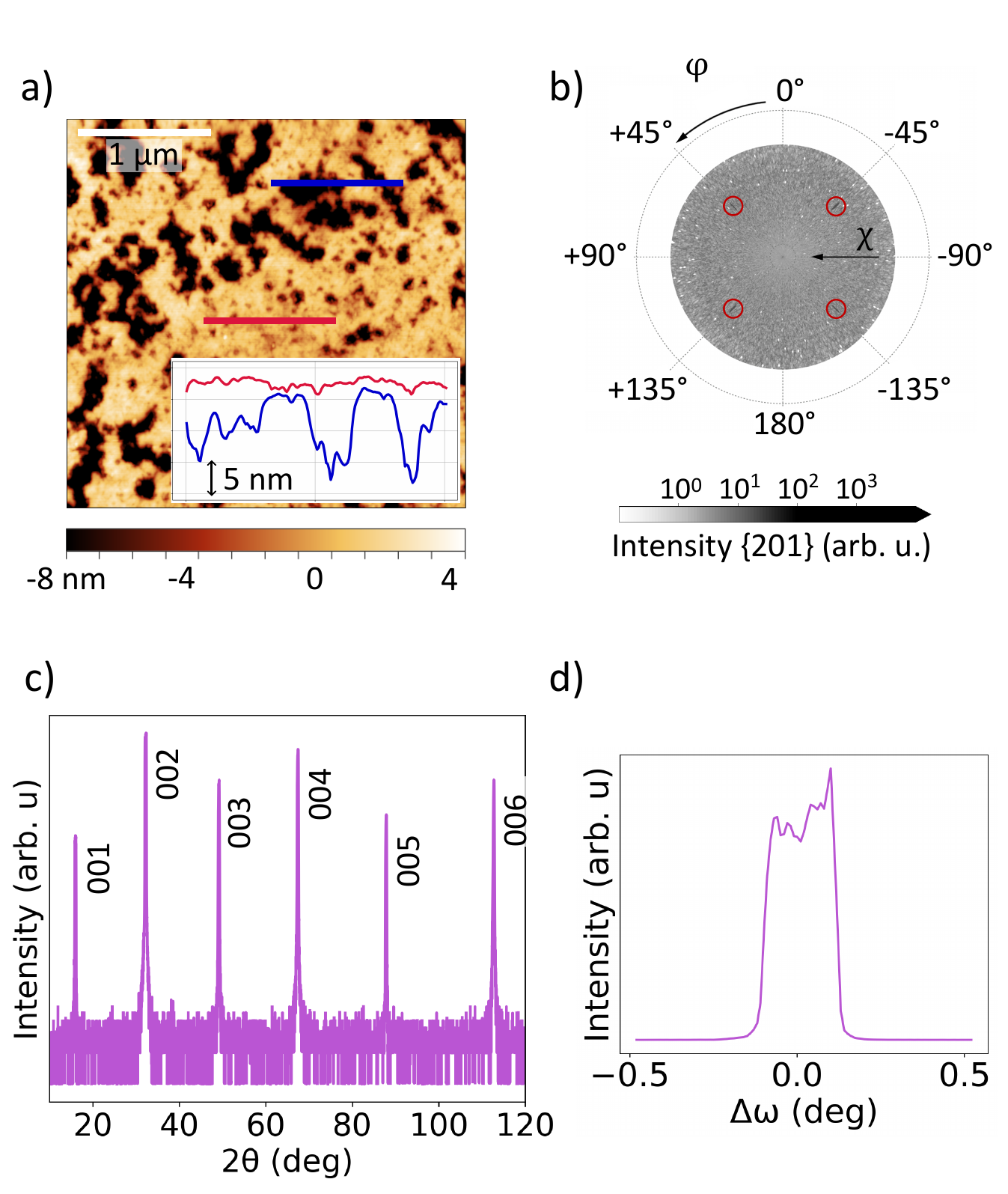}
\caption{(Color Online) a) An atomic force microscopy image of the surface of the BCGO after the chemical-mechanical polishing. The inset shows a height profile along two 1 $\mu$m-long lines, marked in red and blue.  b-d) Structural characterization of the BCGO single crystal at 300 K: b) the $\{201\}$ pole figure, c) a radial scan (the intensity is plotted on logarithmic scale), and d) a rocking curve around (002). Note that the shape of the rocking curve is not a consequence of detector saturation.}
\label{fig:2}
\end{figure}
\begin{figure}[ht!]
\includegraphics[width=0.45\textwidth, trim=0 0 0 0]{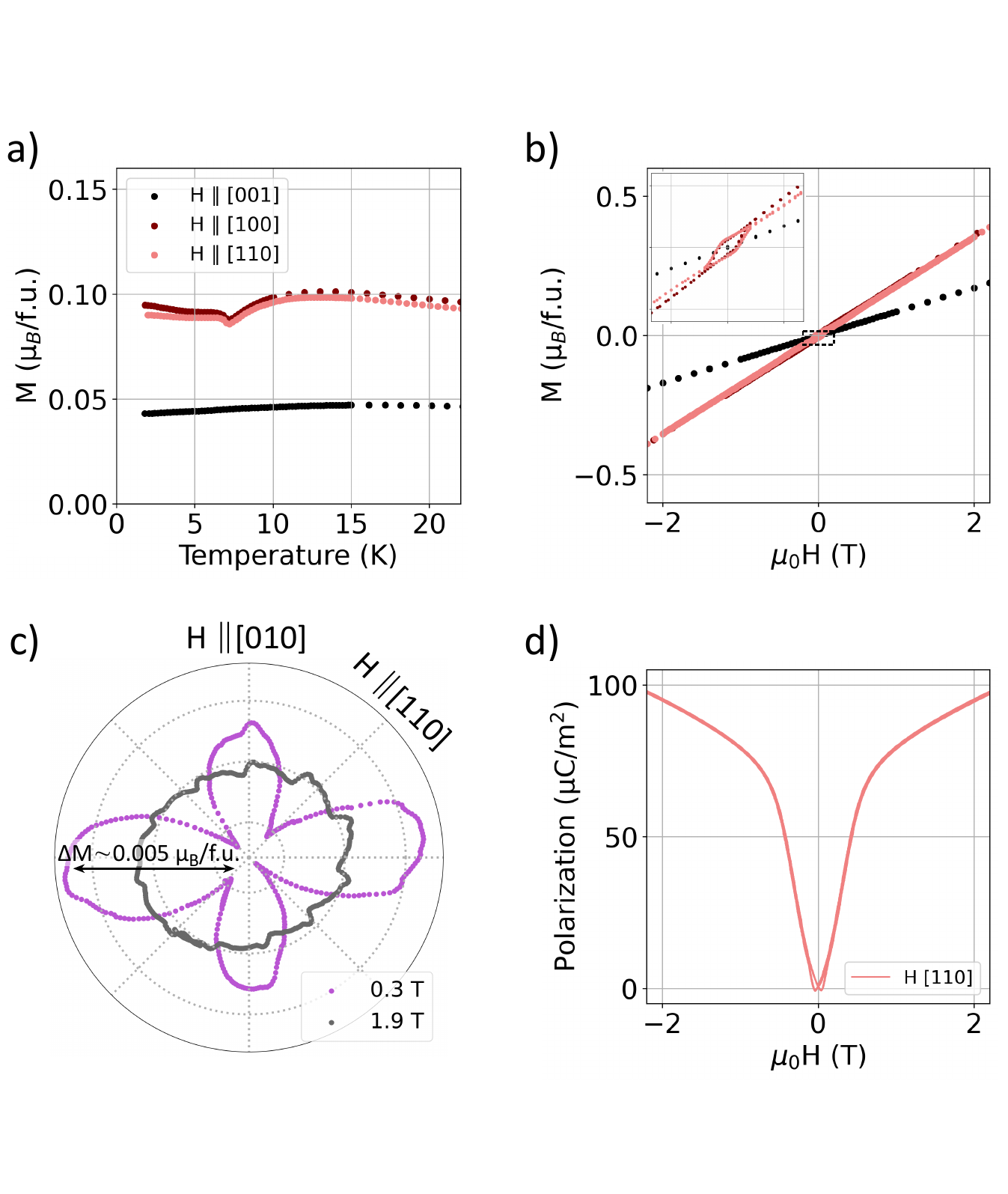}
\caption{(Color Online) a) Temperature-dependence of the magnetization along the external magnetic field measured under a 0.5 T field along different crystal axes. b) Field-dependence of the magnetization measured at 2 K with the field along different crystal axes. The inset is a close-up of the plot marked with the dashed rectangle (in the range 0.2 to -0.2 T and 0.05 to -0.05 $\mu_B$/f.u). See a) for the legend. c) The change in magnetization plotted as $\frac{M-\langle M\rangle}{\langle M\rangle}$, where $\langle M\rangle$ is the average magnetization, as a function of the in-plane field orientation at 2 K. d) The magnetic field-dependence of the polarization along the [001]-direction at 2.5 K when the field is applied along the [110]-direction.}
\label{fig:3}
\end{figure}
Measurements were performed in quantitative nanomechanical mapping intermittent regime and processed using the Gwyddion software \cite{Neas2011}. The BCGO surface is shown in Figure \ref{fig:2}a - the global root mean square (RMS) roughness is $\sim$ 5 nm because about 20 \% of the surface is covered by deeper holes each with an area of the order of 0.1 $\mu$m$^2$ and depth ranging from 5-10 nm. The smoother areas around the holes have $\sim$ 1 nm RMS roughness. X-ray diffraction measurements, shown in Figure \ref{fig:2}b-d, were carried out to confirm the crystallographic structure and orientation of our BCGO bulk sample with a Rigaku Smartlab diffractometer in the parallel beam configuration (using Cu-K$\alpha_1$ radiation). A pole figure was measured for the $\{201\}$ planes to confirm the single-crystalline nature and the four-fold rotational symmetry of the tetragonal crystal structure as shown in Figure \ref{fig:2}b. Four $\{201\}$ poles (or diffraction maxima) are observed with an azimuthal periodicity of 90° as expected. The relevant in-plane crystallographic directions were determined based on the azimuthal dependence. A symmetric radial scan (2$\theta$/$\omega$) measured with this alignment is presented in Figure \ref{fig:2}c. Using these peak positions and a reciprocal space map, lattice constants of a = b = 8.39 \AA, c = 5.55 $\mathrm{\AA}$ were extracted, which are in excellent agreement with the values reported earlier for BCGO \cite{Hutanu2011}. Both the pole figure and the rocking curve of the (002) Bragg reflection (see Figure \ref{fig:2}d) demonstrate that the (001) lattice planes are parallel to the sample surface. We have also measured the chemical composition of the bulk crystals using X-ray fluorescence and the obtained stoichiometry of Ba$_{2.03}$Co$_{1.01}$Ge$_{1.96}$O$_{7}$ is very close to the nominal one of Ba$_{2}$CoGe$_{2}$O$_{7}$.

We have confirmed the very small magnetization in our samples by SQUID magnetometry as shown in Figure \ref{fig:3}a-c, compatible with a compensated magnetic order with slightly canted moments in-plane. The magnetization as a function of temperature, field, and field orientation was measured using the horizontal rotator option (MPMS3, Quantum Design). First, in Figure \ref{fig:3}a, we have identified the critical temperature as $\sim 7$ K, which is in good agreement with the previously reported value of 6.7 K \cite{Sato2003, Yi2008}. As shown in Figure \ref{fig:3}b, we observe a small spontaneous net moment of $\sim$ 0.01 $\mu_B$/f.u. along the [100] and [110] directions, and vanishing spontaneous moment in the [001] direction, which is in agreement with the easy-plane anisotropy as well as DMI-allowed moment canting in the (001) plane \cite{Sato2003, Solovyev2015, Thoma2022}. Considering the slightly larger magnetization when H $\parallel$ [100] (sublattice magnetic moments perpendicular to H) seen in field-sweeps and the in-plane field orientation scans shown in Figure \ref{fig:3}b and c, respectively, we also identify $\langle$100$\rangle$ as the slightly easier axes of the sublattice moments. Note that as shown in Figure \ref{fig:3}c this anisotropy is not present at 1.9 T, which is the external magnetic field strength used for the SMR measurements later on. 

Moreover, we have confirmed the ferroelectric nature of our sample by measuring the polarization along [001], as shown in Figure \ref{fig:3}d. The ferroelectric polarization was measured on a thin slab of BCGO using an electrometer in charge measurement mode (6517A, Keithley) and a cryostat (PPMS, Quantum Design). A finite polarization along the [001] axis is observed when the spins align along the oxygen-oxygen bonds of the CoO$_4$ tetrahedra - upon 90 degree rotation of the spins this polarization reverses the sign whereas upon 45 degree rotation it vanishes  \cite{Murakawa2010}. Applying a magnetic field along the [110] axis will align the spins along the oxygen-oxygen bond and thus yield a finite polarization along the [001] axis, as we see in Figure \ref{fig:3}d. This is in good agreement with previous reports \cite{Murakawa2010}. Removing the magnetic field returns the system to a multidomain state where an equal population of the four magnetic domains with mutually orthogonal N\'eel vector orientations \cite{Vit2021} have vanishing polarization.

\begin{figure}[ht!]
\includegraphics[width=0.49\textwidth, trim=0 0 0 0]{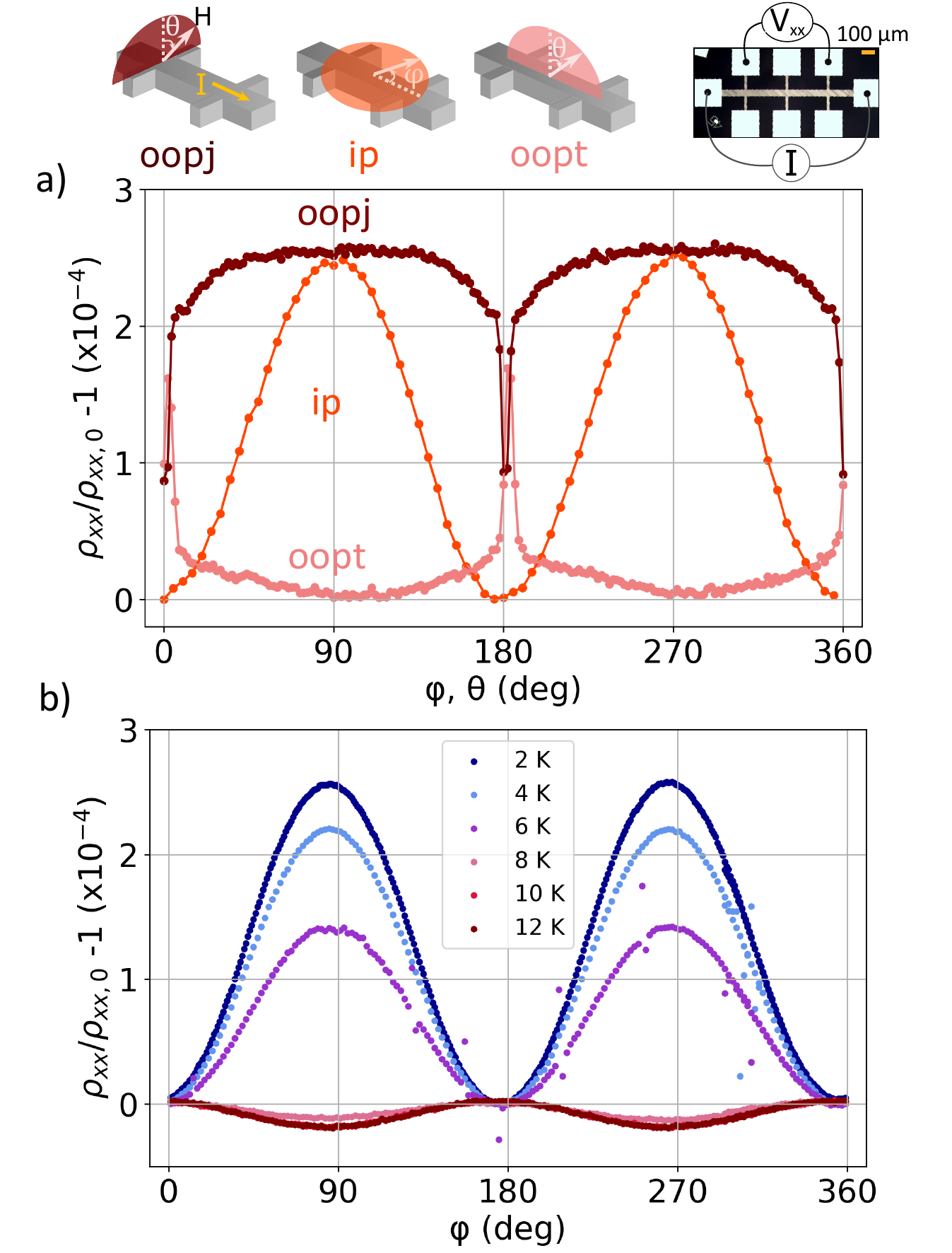}
\caption{(Color Online) a) The SMR signal in BCGO/Pt heterostructure at 2 K for $\mathbf{j_c}\parallel[100]$. A 1.9 T field is rotated in three perpendicular planes (ip, oopj, oopt). The top left inset shows the definitions of the field rotation planes and the top right inset is an optical image of the Hall bar devices. b) The temperature dependence of the SMR for ip-rotation of 1.9 T field. The positive SMR vanishes above the critical temperature of 6.7 K of BCGO.}
\label{fig:4}
\end{figure}

For the SMR measurements, Hall bar structures (width $=50$ $\mu$m, length $=600$ $\mu$m) with different orientations of the current direction with respect to the BCGO [1$\bar{1}$0] crystal axis (defined by angle $\alpha$) were patterned on the polished BCGO (001) surface with e-beam lithography followed by a deposition of 15 nm of Pt through sputtering (Pt resistivity is of the order of 80 $\mu\Omega$cm). After this, the contact pads were patterned by depositing Ti(5)/Au(80 nm) and a subsequent lift-off process. The final devices are shown in the inset of Figure \ref{fig:4}a. To measure the SMR signal, we rotated an external magnetic field of 1.9 T in three mutually orthogonal planes - in-plane (ip), out-of-plane perpendicular to the current $j$ (oopj), and out-of-plane parallel to $j$, (oopt) - while measuring the longitudinal Pt resistivity ($\rho_{xx}$) at a temperature of 2 K. At 2 K we are well below the critical magnetic ordering temperature, and at 1.9 T well above the in-plane monodomainization field and the in-plane magnetocrystalline anisotropy field \cite{Hutanu2012, Hutanu2014} so we assume a smooth rotation of the magnetic order parameter in-plane. The resistivity of bare BCGO is larger than 10$^{17}$ $\Omega$cm at 2 K, confining the current to the Pt layer and therefore excluding contributions from the magnetoresistance of BCGO in the measured SMR signal. We also rule out contributions from proximity-induced magnetic moments in Pt, as the SMR signal persists in control samples with a Cu-spacer between the BCGO and Pt, which decouples the two layers magnetically \cite{SM, Nakayama2013}.

We first focus on the results obtained for a Hall bar with the current direction aligned with the BCGO [100]-direction ($\alpha=$ 45 deg). The modulation of $\rho_{xx}$ during the three field rotations is shown in Figure \ref{fig:4}a (plotted as $\rho_{xx}/\rho_{xx,0}-1$, where $\rho_{xx,0}$ is the Pt resistivity when the magnetic field is parallel to the current, i.e. $\varphi$ = 0). In the in-plane scans we observe a typical SMR response of a $-\cos2\varphi$, where $\varphi$ is the angle between the magnetic field and the current. This behavior is the same as in other systems where the sublattice moments align perpendicular to the magnetic field \cite{Han2014, Hoogeboom2017, Fischer2018, Fischer2020},
indicating that the signal is governed by the sublattice moment orientation rather than that of the weak net magnetization. The temperature-dependence of the SMR signal shown in Figure \ref{fig:4}b matches well with the temperature dependent magnetic ordering of BCGO as shown in Figure \ref{fig:3}a. The positive SMR signal vanishes above the critical temperature indicating that the signal is related to the magnetic order of BCGO. The small negative SMR signal ($\sim-1\times10^{-5}$) right above the critical temperature may originate from the paramagnetic moments aligning with the external field \cite{Schlitz2018, Lammel2019, Oyanagi2021}. This is supported by the fact that we observe this signal also in the BCGO/Cu/Pt control sample \cite{SM} meaning that we can exclude contributions from the proximity-effect induced magnetic moments in Pt.

In the two out-of-plane scans we observe behavior similar to that in $\alpha$-Fe$_2$O$_3$ \cite{Fischer2020}: in the oopj (oopt) planes the SMR exhibits plateaus at maximum (minimum) SMR, respectively, with deep dips (peaks). This indicates that the system likely breaks into a multidomain state when the field is out-of-plane (dips/peaks)
\begin{figure}[ht!]
\includegraphics[width=0.49\textwidth, trim=0 0 0 0]{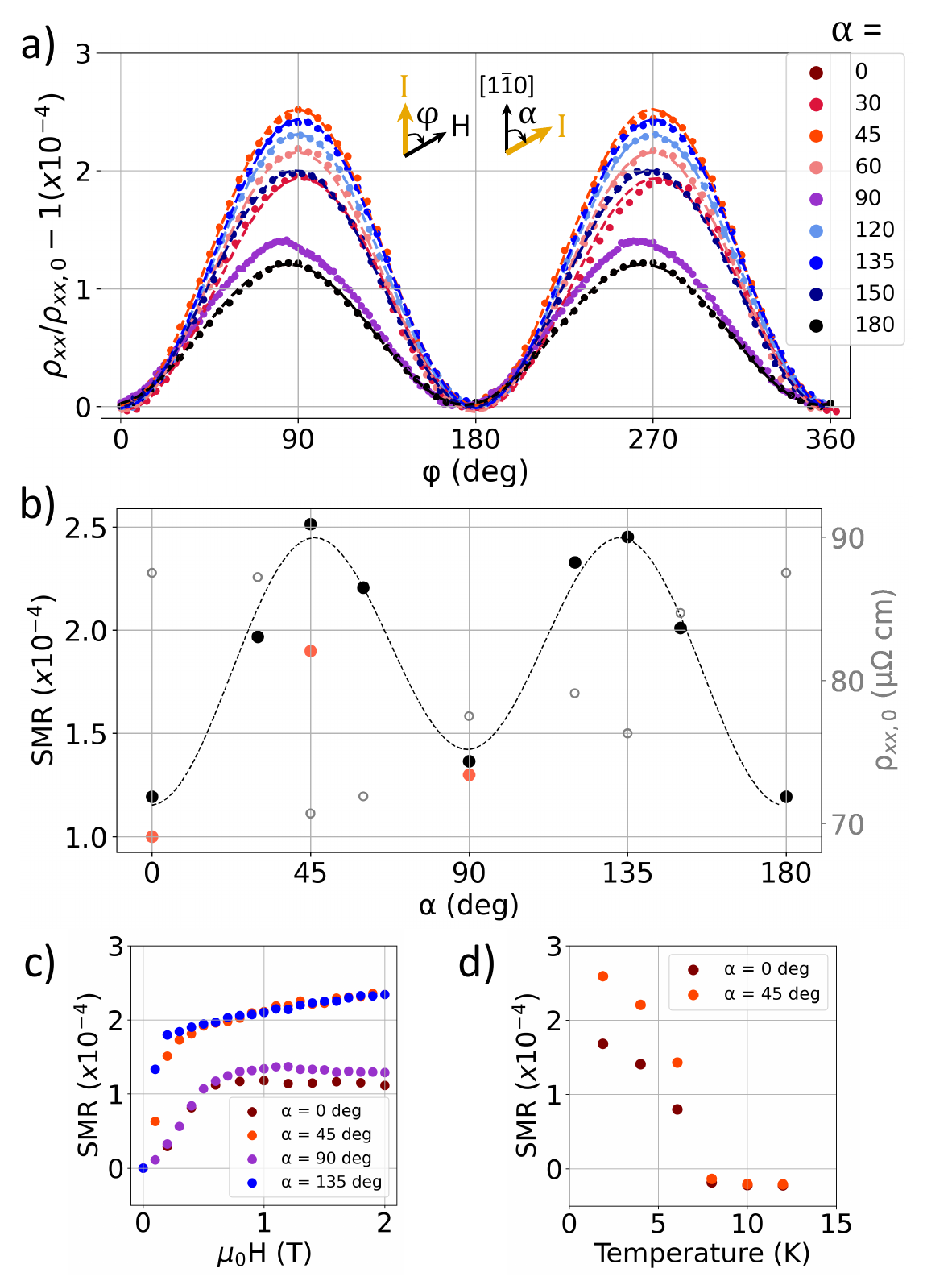}
\caption{(Color Online) a) The full ip angular scans of the SMR for different orientations of the current channel. The definitions for angles $\varphi$ and $\alpha$ are shown in the inset. The measurements were conducted at 2 K and under 1.9 T field. Note that the data for $\alpha =$ 0 and 180 deg is identical. b) The dependence of the SMR ratio (closed symbols) and the longitudinal resistivity (open symbols) at $\varphi=0$ deg on the current channel orientation $\alpha$. The orange closed symbols represent the data measured for the independently fabricated and measured reference sample at 4 K and 1 T.} c) The field-dependence of the SMR ratio for $\alpha=$ 0, 45, 90, and 135 deg at 2 K. The slight amplitude variations compared to a) likely result from slight misalignment of the field orientation from perfectly parallel/perpendicular to the current channel. d) The temperature-dependence of the SMR ratio for $\alpha=$ 0 and 45 deg under 1.9 T field.
\label{fig:5}
\end{figure}
while with an increasing in-plane component of the field the system transitions towards a monodomain state (plateaus). BCGO has two mutually orthogonal pairs of 180 $\deg$ magnetic domains that are degenerate when the field is along [001] direction \cite{Hutanu2014, Vit2021}, which supports the multidomain origin of the dips/peaks. However, compared to $\alpha$-Fe$_2$O$_3$ \cite{Fischer2020}, the plateaus are narrower and the peaks/dips are wider, which can be a result of different domain dynamics and magnetocrystalline anisotropies. Considering the easy-plane magnetocrystalline anisotropy and the argument that an external field along [001] should result in equal population of the domains by symmetry, this domain evolution is a plausible explanation. For a rigorous experimental confirmation of this hypothesis, mapping of the BCGO domain evolution upon external field rotation is required.

We note that the SMR ratio of $\sim$2.5x10$^{-4}$ is surprisingly large considering that the Pt has been deposited \textit{ex situ} without any treatment of the interface to decrease the roughness (see Figure \ref{fig:2}a) or to remove possible impurities, the Pt thickness here is much larger than the spin diffusion length of Pt, and the temperature is very low. The SMR ratio is already comparable to heterostructures with Pt interfaced with conventional antiferromagnets such as NiO with \textit{ex situ} grown Pt with optimized thickness \cite{Hoogeboom2017}. We expect that improving the BCGO surface quality in terms of impurities and crystallinity can further increase the SMR ratio, as has been shown extensively in the case of yttrium ion garnet \cite{Burrowes2012,Jungfleisch2013,Qiu2013,Qiu2015,Putter2017}. This highlights the question whether altermagnetic insulators show larger SMR ratio compared to their ferromagnetic and antiferromagnetic insulator counterparts.

Next, we measured the SMR ratio in Hall bars with different current channel orientations $\alpha$. In Figure \ref{fig:5}, we show that the SMR ratio depends systematically on the crystal direction of the current: when the current flows along [1$\bar{1}$0] or [110] ($\alpha=$ 0 or 90 deg, respectively), the SMR ratio is about a factor of two smaller than the SMR ratio when the current flows along [100] or [010] ($\alpha=$ 45 or 135 deg, respectively). This trend does not correlate with the Pt resistivity of each Hall bar at 2 K as confirmed in Figure \ref{fig:5}b, indicating that the change in the SMR ratio is unlikely to stem purely from device-to-device variations. This is also corroborated by our observation of the same anisotropic behavior in a separate, independently fabricated sample, shown in Figure \ref{fig:5}b. The reference sample has a thinner Pt layer ($\sim$ 8 nm) and it has been measured at 4 K and under 1 T external magnetic field (as opposed to the main device  2 K and 2 T of the main device) resulting in slight discrepancies in SMR amplitudes but nevertheless the symmetry of the anisotropy is well preserved. This highlights the robustness of the current-direction anisotropy of the SMR amplitude.

Next, to exclude different multidomain states as an explanation for the different SMR ratios between $\alpha=$ 0 and 90, and $\alpha=$ 45 and 135 deg devices, we have measured the field dependence of the SMR ratio. To that end, we have measured the field dependence of $\rho_{xx}$ at $\varphi=$ 0 and 90 deg: for $\alpha=$ 0 and 90 deg this means with the field along [1$\bar{1}$0] and [110], and for $\alpha=$ 45 and 135 deg  with the field along [100] and [010]. The resulting field-dependence of the SMR ratio is plotted in Figure \ref{fig:5}c. In the case of each $\alpha$, (relative) saturation is reached well below 1.9 T, suggesting a monodomain state in the in-plane field rotation measurements in Figure \ref{fig:5}a. The magnetocrystalline anisotropy observed in Figure \ref{fig:3}c is also reflected in the slightly different saturation fields between $\alpha=$ 0 and 90 deg and $\alpha=$ 45 and 135 deg. Finally, the anisotropy disappears at 7 K along with the magnetic ordering of BCGO, as shown in Figure \ref{fig:5}d. 

We have also considered the possible influence of the electric polarization along the [001] axis ($P_c$) on the SMR ratios. If we define angle $\alpha'$ as the angle between the external magnetic field and the [1$\bar{1}$0] crystal axis, then $P_c$ has roughly a $-\cos2\alpha'$ dependence, i.e. $P_c$ is non-zero when the magnetic field is along $\langle 110 \rangle$ direction and vanishes when the field is along $\langle 100 \rangle$ direction \cite{Murakawa2010}. On the other hand, the angle $\varphi$ in Figure \ref{fig:5}a is defined as the angle between the current and the field so that for the different Hall bars, non-zero $P_c$ is reached at different angles $\varphi$. If the non-zero $P_c$ influenced the SMR amplitude we would expect to observe this influence at different $\varphi$ for the different $\alpha$. In this case, we consider that observing a $-\cos2\varphi$ behavior for all $\alpha$ would be unlikely and rather we would expect a more complex signal coming from the superposition of the $-\cos2\varphi$ (conventional SMR) and $\cos2\alpha'$ (influence of the $P_c$) behaviors. Another contribution we must consider is the magnetoelectric coupling, as an electric field along [100] direction can induce a finite electric polarization along [100], which then rotates the sublattice magnetic moments away from the (001) plane \cite{Murakawa2010}. However, this requires strong applied electric fields of the order of 1 MV/m while the internal electric fields in the Pt Hall bars will be of the order of 100 V/m. Therefore, we consider the influence of such magnetoelectric coupling negligible in our SMR measurements.

\section{Discussion}

As discussed above, we cannot explain the observed current orientation dependence of the SMR ratio in BCGO/Pt heterostructures with magnetocrystalline anisotropy, multidomain effects, or polarization-induced changes as our SMR measurements are conducted under magnetic fields sufficient to saturate the SMR ratio and the symmetry of the variation in polarization is not compatible with that of the SMR ratio. We also note that the Hall bars measured here are all fabricated on the same surface. This ensures that - unlike in the case of anisotropic SMR observed in ferrimagnetic insulators \cite{Isasa2014, Wu2021} - any crystal cut-related variations in spin mixing conductance, such as differences in surface roughness or variations in magnetic ion area density do not contribute to the observed anisotropy. 

Conversely, if any of the three spin current channels contributing to SMR (magnon excitation, magnon capacitance, and STT, as shown in Figure \ref{fig:1}b) exhibit anisotropy with respect to the current direction, this is would be reflected in the anisotropy of the SMR ratio. It has been previously shown that the anisotropic local crystal environment inherent to altermagnets naturally leads to magnetotransport properties that exhibit unconventional anisotropy with respect to the crystal direction \cite{Bose2022, Feng2022, Bai2022, Karube2022, Bai2023, GonzalezBetancourt2024, Leiviska2024, GonzalezBetancourt2024}. This scenario is, in principle, plausible for several reasons: first, both the thermal magnon excitation \cite{Cornelissen2016, Reiss2021} and STT contributions \cite{Chen2013, Reiss2021} depend on the spin mixing conductance, which is predicted to vary with crystal orientation due to crystal field effects \cite{Cahaya2017}. Second, the magnon-related contributions depend on the magnon density of states, which is influenced by the magnon dispersion relation \cite{Kato2020, Reiss2021}. The magnon density of states can exhibit anisotropy with respect to crystal momentum due to magnetocrystalline anisotropy (a relativistic effect) or due to anisotropic exchange interactions (a non-relativistic effect)  \cite{Smejkal2023, Cui2023}. Finally, the magnon capacitance contribution, which depends on the exchange interactions and the magnon lifetime \cite{Reiss2021, Franke2024}, may also exhibit anisotropy with
crystal momentum.

BCGO is a planar g-wave altermagnet \cite{Smejkal2024}, i.e. it exhibits four nodal planes in the (001) plane. We have confirmed the g-wave symmetry using DFT calculations \cite{SM}. Our calculations show that the non-relativistic spin splitting in BCGO is very small and therefore unlikely to explain the anisotropic SMR we observe in Figure \ref{fig:5}b. Instead, the anisotropy (that can be fitted with a $-\cos2\alpha-\cos4\alpha$ dependence) is more compatible with the relativistic magnetic space group that is orthorhombic: when the magnetic moments point along $\langle 110 \rangle$ axes, the space group is Cm’m2’ while when the moments point along $\langle 100 \rangle$ the space group is P2$_1$2$_1$’2’. We have calculated also the relativistic electronic band structure \cite{SM} that confirms the orthorhombic symmetry. In order to map the non-relativistic and relativistic symmetries of BCGO to the anisotropy of the overall SMR ratio, it is important to note that the latter represents a superposition of the anisotropy of each spin current channel. Furthermore, anisotropy in one physical parameter such as the spin mixing conductance or magnon density of states can influence more than one spin current channel. Rigorous confirmation of the origin of the observed current-direction anisotropic SMR therefore requires additional systematic and extensive theoretical modeling and symmetry analysis of the anisotropy in the relevant physical parameters such as the spin mixing conductance and magnon density of states. While this is beyond the scope of the present paper, it represents an important future task. Experimentally, the origin of the anisotropic SMR ratio could be studied further by extending SMR measurements beyond the dc limit into the THz range and thus leveraging the different frequency-dependencies of the different spin current channels \cite{Reiss2021} to disentangle their contributions to the overall SMR ratio. In addition to providing further information on the anisotropy of the different terms, such experiments could allow identifying whether any of the terms can explain the surprisingly large SMR ratio reported in the altermagnetic MIs (this work and Ref. \cite{Fischer2020}) compared to their ferromagnetic and antiferromagnetic counterparts.

Finally, we will address whether the anisotropic SMR is a unique feature of BCGO or can extend to other altermagnetic candidates. For a systematic comparison, we consider only insulators (where the electronic SMR channels are not active \cite{Reiss2021}) and single crystals (where all the magnonic contributions can be expected to be present \cite{Reiss2021}). SMR ratios for two mutually perpendicular Hall bars have been reported for single crystals of $\alpha$-Fe$_2$O$_3$ \cite{Lebrun2019} (another altermagnetic candidate material \cite{Smejkal2022a, Verbeek2024, Galindezruales2024}), indicating anisotropy with the crystal-direction of the current channel. However, the origin of this anisotropy is primarily discussed in terms of the N\'eel vector dynamics governed by the magnetocrystalline anisotropy while systematic studies on more Hall bar orientations are still lacking. In order to conclusively address the universality of the SMR ratio anisotropy with the current direction in altermagnetic insulating bulk MI layers, further systematic experiments are required on materials such as $\alpha$-Fe$_2$O$_3$, and will be an important future research direction. 

\section{Conclusion}

In this work, we have shown that the heterostructure of Pt and a single crystal of the insulating and multiferroic altermagnetic candidate Ba$_2$CoGe$_2$O$_7$ exhibit a surprisingly large, current-direction anisotropic spin Hall magnetoresistance signal. We have discussed the anisotropy in the context of magnetocrystalline anisotropy, magnetic domains, and variation in the electric polarization and concluded that these factors are unlikely to provide a complete explanation. Finally, we have discussed the possible mechanisms through which the non-relativistic and relativistic symmetries of Ba$_2$CoGe$_2$O$_7$ could give rise to the current-direction anisotropic SMR. To unambiguously understand the origin of the observed anisotropy, a quantitative theoretical model that takes into account the possible anisotropy in all the various contributions to SMR as well as further systematic experimental SMR studies - in particular in the THz regime - on heterostructures with a heavy metal and an altermagnetic insulator are required. Overall, our findings represent an initial step toward a deeper understanding of the current-direction anisotropic spin Hall magnetoresistance in single crystalline altermagnetic candidate materials as well as toward the realization of devices with tunable and electrically controllable SMR response.

\nocite{Ceperley1980,Kresse1999, Kresse1996a, Kresse1996b}

\begin{acknowledgments}

M. L. is co-funded by the European Union (Physics for Future – Grant Agreement No. 101081515). The work was also supported by the Grant Agency of the Czech Republic Grant No. 22-17899K, the Dioscuri Program LV23025 funded by MPG and MEYS, TERAFIT - CZ.02.01.01/00/22$\_$008/0004594 funded by OP JAK, call Excellent Research, the Deutsche Forschungsgemeinschaft (DFG, German Research Foundation) via projects 445976410 and 490730630, and the SFB 1432, Project-ID No. 425217212, and the Grant Agency of the Charles University through grants No. 166123 and SVV–2024–260720. We also gratefully acknowledge technical support and advice by the nano.lab facility of the University of Konstanz. V. K. was supported by the Alexander von Humboldt Foundation. AUBW and BB were supported by DFG through SFB 1143, project-id 247310070 and the Würzburg-Dresden Cluster of Excellence on Complexity and Topology in Quantum Matter ct.qmat (EXC 2147, project-id 390858490)) This work was also supported by CzechNanoLab project LM2023051 funded by MEYS CR. DK acknowledges the Lumina Quaeruntur fellowship LQ100102201 of the Czech Academy of Sciences.
\end{acknowledgments}

\newpage

\bibliography{biblio.bib}
\end{document}